\documentclass{article}
\usepackage{amsthm, amssymb}
\newcommand{\abs}[1]{\vert #1 \vert}
\newcommand{\gen}[1]{\langle #1 \rangle}
\newcommand{\ket}[1]{\vert #1 \rangle}
\newtheorem{theorem}{Theorem}
\newtheorem{proposition}[theorem]{Proposition}
\newtheorem{definition}[theorem]{Definition}

\begin{document}

\title{Quantum Property Testing for Solvable Groups}
\author{Yoshifumi Inui\thanks{psi@is.s.u-tokyo.ac.jp}\and\small{Department of Computer Science, The University of Tokyo}\and\small{ERATO-SORST Quantum Computation and Information Project, JST}}


\date{}

\maketitle

\begin{abstract}
Property testing has been extensively studied and its target is to determine whether a given object satisfies a certain property or it is far from the property.
In this paper, we construct an efficient quantum algorithm which tests if a given quantum oracle performs the group multiplication of a solvable group.
Our work is strongly based on the efficient classical testing algorithm for Abelian groups proposed by Friedl, Ivanyos and Santha.
Since every Abelian group is a solvable group, our result is in a sense a generalization of their result.
\end{abstract}


\section{Introduction}
In order to guarantee the behavior of a program, we must prove it mathematically.
For an ideal algorithm, constructing a perfect proof can be done, but for practical software, it is usually difficult and tedious.
Furthermore such an approach cannot detect hardware based errors (common in the quantum setting).
Testing the outputs through all inputs ensures the correctness of an algorithm exactly but it takes too much time.
In this context, program checking \cite{blum,bk}, self-testing \cite{blr} and self-correcting \cite{blr,lipton} were introduced.
A program checker checks if a program is correct on a particular input.
A self-tester tests if a program is correct on most inputs.
A self-corrector computes a correct value using a program which is correct on most inputs.
In the study of these methods, the relation between a function $f$ itself and a program $P$ expected to compute $f$ matters.

Property testing is another checking technique.
It deals with efficient algorithms deciding whether a given object has some expected property or it is far from any object having that property.
In testing a function, the difference between property testing and self-testing is that
property tester tests only if a program has the property which the function has, 
and does not test if a program's behavior is the function itself.
One of the interests of property testing is that, by running property testing algorithm in advance self-testing tasks become quite easy in many cases.
The notion of property testing was first introduced by Rubinfeld and Sudan \cite{rubinfeld}.
In this setting many properties including algebraic function properties \cite{rubinfeld,ergun,ben-or,friedl}, 
graph properties \cite{goldreich,afns}, computational geometry properties \cite{adpr,czumaj00} and regular languages \cite{anks,flnrrs,fkrss} were proved to be testable.
Quantum testers have also been studied and they are known to be more powerful than classical testers \cite{buhrman,fmss,magniez}.
See \cite{fischer,ron} for surveys on property testing.

The most active research areas in quantum computing are group theoretic problems.
Shor's factorization algorithm \cite{shor} can be regarded as an instance of the hidden subgroup problem (HSP) (see \cite{lomont} for a survey).
The HSP over dihedral groups and symmetric groups are interesting and challenging
because the former corresponds to the unique shortest vector problem \cite{regev} and the latter to the graph isomorphism problem.
Other quantum algorithms for group theoretic problems \cite{watrous,fenner} are also known.
In most of these problems, group operations are carried out using group oracles.
Given two elements $a, b$ in a group, the group oracle returns $a\cdot b$, 
and it returns the inverse $a^{-1}$ given an element $a$.
However its structure is unknown to us in the setting.
Thus a natural question arises: is a given group oracle really what we expect?

Now our concern is to test if a given oracle has the property as a supposed group oracle.
This test is related to the tests to decide whether a given oracle is that of some group and whether a function from a group to another group is a homomorphism. 
The second test has been studied \cite{blr,ben-or} and it can be done efficiently in the classical setting.
About the first test, classical algorithms testing associativity of given functions were constructed in \cite{ergun,rajagopalan}.
These algorithms can be used for general groups, but their running times are polynomial in the size of the ground sets.
As each elements in a set $\Gamma$ is encoded in $O(\log\abs\Gamma)$ bits, time complexities are exponential of the input lengths.
However, Friedl, Ivanyos, and Santha have succeeded in designing a polynomial time algorithm which tests whether a given oracle is an Abelian group oracle or not \cite{friedl}.
Their testing algorithm is a classical one, but its basic idea came from the quantum setting.

In this paper, we construct an efficient quantum algorithm for the problem of testing whether a given oracle is that of a solvable group.
The definition of solvable group is as follows.
\begin{definition}
\label{def-sg}
A group $G$ is solvable if it has a normal subgroup chain $\{1\}=G_0\lhd\cdots\lhd G_t=G$
such that $G_{i+1}/G_i$ is an Abelian group for all $i$.
\end{definition}
It is obvious that an Abelian group is a solvable group.
We can find much more solvable groups in quantum computation areas.
Dihedral groups over which the HSP is solvable in subexponential time \cite{kuperberg} are solvable groups.
Moreover some non-Abelian groups over which the HSP are efficiently solvable are solvable \cite{rotteler,hallgren,fimss,moore,psi,bacon}.
Watrous's algorithm \cite{watrous} and Fenner and Zhang's algorithm \cite{fenner} are also for solvable groups.
As can be seen from the above examples, quantum oracles for solvable groups are used in many important algorithms.
Hence it is natural to focus on the problem of testing if given oracles are really solvable group oracles.

\section{Definitions}
We give a formal definition for property testers.
\begin{definition}
Let $f$ be a function, $P$ a set of functions, and $d$ a distance function between functions.
A quantum $\epsilon$-tester for $P$ with distance $d$ is a quantum oracle Turing machine $M$ such that
$$\left\{\begin{array}{l l l}
	{\bf Pr}[M^f(\epsilon)\, accepts]>\frac 2 3\quad if\, d(f, P)=0\\
	{\bf Pr}[M^f(\epsilon)\, accepts]<\frac 1 3\quad if\, d(f, P)>\epsilon.
\end{array}\right.$$
We use $d(f, P)$ to represent $\inf_{g\in P}d(f,g)$ here.
\end{definition}

Just like the Abelian case, in this paper we adopt the edit distance as the distance function, which we now define.
The complete definition is described in \cite{friedl}.
A table is a square matrix and its size is the number of elements.
We consider three operations to transform a table to another.

An exchange operation replaces elements in a table by arbitrary elements and its cost is a number of replaced elements.
An insert operation inserts new rows and columns.
A delete operation deletes existing rows and columns.
The indices of inserted/deleted rows and columns must be equivalent. 
Its cost is a number of inserted/deleted elements.

We define edit distance for tables using the idea of transform, and extend it to magmas.
A magma $M$ is a set equipped with a binary operation $\cdot:M\times M\rightarrow M$.
A multiplication table of a magma $M$ is a table of which both rows and columns have a one-to-one correspondence with elements in $M$,
$\phi:\{1,\cdots,\abs M\}\rightarrow M$, and elements in the $i$-th row and the $j$-th column is equivalent to $\phi(i)\cdot\phi(j)$.
\begin{definition}
Edit distance between two tables $T$ and $T'$ is the minimum cost needed to transform $T$ to $T'$ divided by the maximum size between $T$ and $T'$.

Edit distance between two magmas $M$ and $M'$ is the minimum edit distance between $T$ and $T'$
where $T[T']$ runs over all tables which is a multiplication table of $M[M']$.
We identify two multiplication tables if we can obtain one from another only by renaming elements in its ground sets.

We say that a magma $M$ is $\epsilon$-close to another magma $M'$ if the distance between them is less than or equal to $\epsilon$,
and that M is $\epsilon$-far from $M'$ otherwise.
\end{definition}

\section{Algorithm}
\subsection{Overview}

We are given a set $\Gamma$ and a quantum oracle which performs a binary operation $\cdot$ over $\Gamma$.
This quantum oracle carries out the following unitary operations: $\ket a\ket b\mapsto\ket a\ket{a\cdot b}$ for all $a, b\in\Gamma$.
\begin{theorem}
There exists a quantum $\epsilon$-tester for solvable groups which runs in polynomial time in $\log\abs\Gamma$ and $\epsilon^{-1}$.
\end{theorem}
Here, we identify a magma with its multiplication function.

We now overview our testing algorithm.
The details are given in subsequent subsections.

First, we pick sufficiently many random elements $\alpha_1, \cdots, \alpha_s$ from the ground set $\Gamma$.
As in Watrous's algorithm \cite{watrous}, we apply Babai et al.'s polynomial time Monte Carlo algorithm \cite{babai} which tests whether an input group is solvable.
Notice that this algorithm requires that the input is a group,
so we cannot decide if $(\Gamma,\cdot)$ is a solvable group at this stage.
When the input is a group, we obtain the elements $h_1, \cdots, h_t$
satisfying $$\{1\}=G_0\lhd\cdots\lhd G_i=\gen{h_1,\cdots,h_i}\lhd\cdots\lhd G_t=\gen{\alpha_1,\cdots,\alpha_s}.$$

Now we define the set $H_j\subseteq\Gamma$ corresponding to $G_j$, as the following:
$$H_0=\{1\}, H_j=\{h_j^a\cdot h\vert a\in\mathbb Z_{m_j},  h\in H_{j-1}\}.$$
The integer $m_j$ is the order of $h_j$ with respect to $H_{j-1}$, and the way to obtain it is given later.

Since $H_0$ is an identity group, it is obviously a solvable group.
If we can test whether $H_j$ is a solvable group under the assumption that $H_{j-1}$ is a solvable group, 
we can test whether $H_t$ is a solvable group inductively.
We show how to test it.

Assume that $H_{j-1}$ is $\frac\epsilon 2$-close to a solvable group $\tilde H_{j-1}$, multiplication of which can be computed efficiently.
In order to test if $H_j$ is a solvable group, we construct a solvable group which is $\frac\epsilon 2$-close to $H_j$ using $\tilde H_{j-1}$.
For this purpose we use the following theorem also used for Abelian groups.
\begin{theorem}[Theorem 2 of \cite{friedl}]\label{hom}
Let $G$ be a group, $f:G\rightarrow H$ be a function and $\eta<\frac 1{120}$.
Assume that the inequality ${\bf Pr}_{x, y\in G}[f(xy)=f(x)\cdot f(y)]>1-\eta$ holds.
Then there exists a group $\tilde H$ with multiplication $*$ and a homomorphism $\tilde f:G\rightarrow\tilde H $ such that
\begin{enumerate}
\item $\vert\tilde H\backslash H\vert\le30\eta\abs{\tilde H}$,
\item ${\bf Pr}_{\alpha, \beta\in\tilde H}[\alpha*\beta\ne\alpha\cdot\beta]\le91\eta$, 
\item ${\bf Pr}_{x\in G}[\tilde f(x)\ne f(x)]\le30\eta$.
\end{enumerate}
\end{theorem}
We construct a group $G$ and a function $f$ in such a way that $\tilde f$ is an isomorphism and $\tilde H$ is $\epsilon$-close to $H$.
And then for a random pair of $x$ and $y$ we test whether it satisfies $f(xy)=f(x)\cdot f(y)$, which we call a homomorphism test.
We evaluate the probability inequality by conducting homomorphism tests several times.
Applying the theorem, if the inequality is satisfied there exists a solvable group $\tilde H_j$ which is $\frac\epsilon 2$-close to $H_j$.

As a result, we can test whether $H_t$ is a solvable group.
However there might exist elements in $\Gamma\backslash H_t$.
In order to guarantee the fraction of such elements is less than $\frac\epsilon 2$, 
we pick $O(\epsilon^{-1})$ random elements from $\Gamma$ and decompose them over $H_t$.
If there exist elements that cannot be decomposed, we reject.

We summarize the algorithm briefly.

\begin{enumerate}
\item Find generators of a normal subgroup chain.
\item For each $1\le j\le t$ do (a) and (b).
\begin{enumerate}
\item Find the order of $h_j$ w.r.t. $H_{j-1}$.
\item Test if $f$ is a homomorphism.
\end{enumerate}
\item Test if there exist elements in $\Gamma\backslash H_t$.
\end{enumerate}


The rest of this chapter covers the shortcoming of details and has the following organization.
Subsection \ref{ord} deals with the problem of finding the order of $h_j$ with respect to $H_{j-1}$,
which is the key task in Watrous's order finding algorithm for solvable groups \cite{watrous}.
In subsection \ref{decomp}, we show the algorithm to decompose an element over $H_j$.
Utilizing the results obtained in these two subsections,
in subsection \ref{const} we construct a solvable group $G$ and a function $f$ in Theorem \ref{hom}.
With these $G$ and $f$ we apply Theorem \ref{hom} and show that the group $\tilde H_j$ is $\frac\epsilon 2$-close to $H_j$ in subsection \ref{app}.
Correctness and complexity of the algorithm are discussed in subsection \ref{analysis}.

\subsection{The order of $h_j$ with respect to $H_{j-1}$}\label{ord}

When $H_{j}$ is a solvable group, we call $m_j:=\min\{m\vert h_j^m\in H_{j-1}\}$ the order of $h_j$ with respect to $H_{j-1}$,
which can be gained using Watrous's algorithm \cite{watrous}.
Also in a non-group case, we call the value obtained from Watrous's algorithm the order of $h_j$ with respect to $H_{j-1}$.
The details of the algorithm are as follows.

We prepare the superposition state over $\tilde H_{j-1}$, $\ket{\tilde H_{j-1}}=\frac 1{\sqrt{\abs{\tilde H_{j-1}}}}\sum_{h\in\tilde H_{j-1}}\ket h$, 
and the state over $\mathbb Z_r$, $\frac 1{\sqrt r}\sum_{a=0}^{r-1}\ket a$ where $r$ is the order of $h_j$ and can be found by Shor's order finding algorithm \cite{shor}.
Then we apply the following operations:
$$\frac 1{\sqrt r}\sum_{a=0}^{r-1}\ket a\ket{\tilde H_{j-1}}\mapsto\frac 1{\sqrt r}\sum_{a=0}^{r-1}\ket a\ket{h_j^a\cdot\tilde H_{j-1}}
\mapsto\frac 1 r\sum_{a, b=0}^{r-1}e^{2\pi i\frac{ab}r}\ket b\ket{h_j^a\cdot\tilde H_{j-1}}.$$
The latter operation is the quantum Fourier transform modulo $r$.
The multiplication between $h_j^a$ and $\tilde h\in\tilde H_{j-1}$ is $h_j^a\cdot h$ where $h$ is an element in $H_{j-1}$ corresponding to $\tilde h$.
By observing the first register, we obtain $\tilde b$ which is a multiple of $\frac r{m_j}$ since $\ket{h_j^{m_j}\cdot\tilde H_{j-1}}=\ket{\tilde H_{j-1}}$.
Repetition yields the desired value $m_j$.
For such $m_j$ we check $h_j^{m_j}\in H_{j-1}$ by decomposing $h_j^{m_j}$.

\subsection{Decomposition over $H_j$}\label{decomp}

Testing if an element $h$ in $H_j$ is really a member of $H_j$ reduces to the task of decomposing $h$ over $H_j$.
The decomposition algorithm is as follows.

Suppose $h=h_{j-1}^{a_{j-1}}\cdots h_{1}^{a_{1}}$.
First we prepare the following state: $$\sum_{a, b}\ket a\ket b\ket{{h^a}\cdot(h_{j}^b\cdot\tilde H_{j-1})}.$$
We can obtain $a_j$ as we do for discrete logarithm problem,
because $\ket{{h^a}\cdot(h_{j}^b\cdot\tilde H_{j-1})}=\ket{\tilde H_{j-1}}$ for $a$ and $b$ satisfying $h^a\cdot h_j^b\in\tilde H_{j-1}$,
that is $aa_j+b=0$.
Next we do similarly for $h\cdot h_{j}^{-a_{j}}$.
Thus we can find each $a_k$ inductively, and we obtain a decomposition of $h$ finally.

\subsection{Construction of $G$ and $f$}\label{const}

We construct an appropriate solvable group $G$ and a function $f$ in Theorem \ref{hom} and claim that $\tilde H_j$ is $\epsilon$-close to $H_j$ for such $G$ and $f$.

We define the group $G$ first.
We extend $\tilde H_{j-1}$ which is almost isomorphic to $\tilde H_{j-1}$ to a solvable group $\tilde H_j$ by adding $h_j$ so that $\tilde H_{j}$ is almost isomorphic to $H_j$.
Consider a pair of two groups $\mathbb Z_{m_j}\times\tilde H_{j-1}$ and introduce a multiplication $\circ$ over it satisfying
$(m, h)\circ(n, h')=(m+n, \phi^n({h})h')$.
Here, $\phi$ satisfies the following conditions:
$$\left\{\begin{array}{l c l}
	\phi(hh')&=&\phi(h)\phi(h')\\
	\phi(h_i)&=&h_j^{-1}\cdot(h_i\cdot h_j).
\end{array}\right.$$
If $\phi$ is an automorhphism of $H_{j-1}$, $\mathbb Z_{m_j}\times\tilde H_{j-1}$ is a solvable group with the multiplication $\circ$, 
because it satisfies the definition of a group and $(0, \tilde H_{j-1})$ is its normal subgroup.
In order to check if $\phi$ is an automorphism, we must check $\forall i\,\,\phi(h_i)\in\tilde H_{j-1}$ and $\abs{H_{j-1}}=\abs{\phi(H_{j-1})}$.
The former can be done by decomposing $\phi({h_i})$ over $\tilde H_{j-1}$ and the latter by Watrous's algorithm.

We have defined the group $G$, so next define the function $\psi$ like this:
$$\begin{array}{c c c}
\psi:\mathbb Z_{m_j}\times\tilde H_{j-1} &\rightarrow& H_j\\
\quad(n,\tilde h) &\mapsto& h_j^n\cdot h.
\end{array}$$
Here $\tilde h$ is represented as $(a_{j-1}, \cdots, a_{1})$ and $h$ is defined as $h_{j-1}^{a_{j-1}}\cdot\cdots\cdot h_{1}^{a_1}$ for such $a_i$s.

\subsection{Application of Theorem \ref{hom}}\label{app}

In order to conduct a homomorphism test with $G=(\mathbb Z_{m_j}\times\tilde H_{j-1}, \circ)$ and $f=\psi$ which we have constructed in the previous subsection, 
we must compute $\psi(xy)$ for $x, y\in\tilde H_j$ efficiently.
To that end, we need to know the value $\phi^n(h)$ for $h\in\tilde H_{j-1}$.
We show how to obtain it.
Let $h$ be represented as $(a_{j-1}, \cdots, a_{1})$.
Because $$\phi^n({h_{j-1}}^{a_{j-1}}\cdots h_{1}^{a_1})=\phi^n({h_{j-1}})^{a_{j-1}}\cdots\phi^n(h_{1})^{a_{1}}$$
 holds, 
we only need the value $\phi^n({h_i})$ for each $i$ and $n$.
We already know the value $\phi({h_i})$ by decomposition.
Assume we know the value $\phi^{2^k}({h_i})$ for each $i$.
The value $\phi^{2^{k+1}}(h_i)$ is equivalent to $\phi^{2^k}(\phi^{2^k}({h_i}))$, 
hence it can also be computed efficiently.
The value for $n$ which is not a power of two can be computed similarly.
Therefore we can obtain all values inductively and compute the multiplication over $G$ efficiently.

Now by applying the theorem we obtain the homomorphism $\tilde\psi$ and the group $\tilde H_j$,
which we want to conclude is $\frac\epsilon 2$-close to $H_j$.
\begin{proposition}\label{prop}
Assume $H_{j-1}$ is $\frac\epsilon 2$-close to a solvable group $\tilde H_{j-1}$.
If $\psi$ constructed in the previous subsection
satisfies ${\bf Pr}_{x, y\in G}[\psi(x\circ y)=\psi(x)\cdot \psi(y)]>1-\eta$,
the group $\tilde H_j$ the existence of which is guaranteed by Theorem \ref{hom} is $151\eta$-close to $H_j$.
\end{proposition}
\begin{proof}
The distance between $\tilde H_j$ and $H_j$ is determined by the number of elements being a member of either set and
the number of pairs of two elements multiplication of which differ by operation.
In fact, to transform a multiplication table of $\tilde H_j$ to that of $H_j$,
we first delete and insert rows and columns corresponding to elements in $\tilde H_j\backslash H_j$ and $H_j\backslash\tilde H_j$ respectively,
and then exchange multiplication values which differ between two tables.
It follows that the number of elements in $\tilde H_j\backslash H_j$ and the number of pairs,
$\gamma_1$ and $\gamma_2$ such that $\gamma_1\circ\gamma_2\ne\gamma_1\cdot\gamma_2$, are small from 1 and 2 in Theorem \ref{hom}.
To complete the proof, we must show that the size of $H_j\backslash\tilde H_j$ is small.
This follows from the theorem if $\abs{\tilde H_j}$ is greater than or equal to $\abs{H_j}$.

As $\tilde\psi$ is a homomorphism from $G$, $\tilde H_j$ is isomorphic to a subgroup of $G$.
Note that $\tilde H_j$ cannot be isomorphic to a group of the form $(\mathbb Z_{m_j}\times H', \circ)$ where $H'\lneq\tilde H_{j-1}$, 
because the contradiction that unitary operation $\ket{h_j^n}\ket{h}\mapsto\ket{h_j^n}\ket{h_j^n\cdot h}$ is not invertible follows from 3 of Theorem \ref{hom}.
Suppose $\gen{(n_0, \tilde h_0)}$ is the kernel of $\tilde\psi$.
Since it is a subgroup of $G$, it follows that $n_0$ is either $m_j$ or a strict divisor of $m_j$.

Assume $n_0\ne m_j$ holds for contradiction.
Recall how we found $m_j$.
The theorem tell us for almost all $n$ and $\tilde h$ the value $h_j^n\cdot h$ is same with $\tilde\psi(n, \tilde h)$.
Therefore we must have obtained $n_0$ instead of $m_j$ since $\ket{\tilde h_j^{n_0}\cdot\tilde H_{j-1}}=\ket{\tilde H_{j-1}}$.
This is a contradiction.
It follows that $n_0=m_j$, therefore $G\cong\tilde H_j$ holds.
It indicates that $\abs{\tilde H_j}=\abs G=m_j\abs{\tilde H_{j-1}}\ge\abs{H_j}$.
\end{proof}

By taking $\eta$ so that $151\eta\le\frac\epsilon 2$, we can test whether $H_j$ is $\frac\epsilon 2$-close to a solvable group
under the assumption that $H_{j-1}$ is $\frac\epsilon 2$-close to a solvable group $\tilde H_{j-1}$.

\subsection{Analysis}\label{analysis}

In this subsection, we discuss the correctness and the complexity of the algorithm.
We run the above algorithm to test whether a given oracle is that of a solvable group or not.
If an error occurs during the execution we reject, otherwise we accept.
We expect the algorithm to accept a solvable group oracle and reject the others with high probability, 
and work in polynomial time in $\log\abs\Gamma$ and $\epsilon^{-1}$.

It is obvious that a solvable group oracle passes the test with high probability.
We would like to show that the oracle $\epsilon$-far from that of a solvable group is rejected with high probability.
Take an oracle $\epsilon$-far from that of any solvable group, then $H_t$ is $\frac\epsilon 2$-far from $\tilde H_t$ or $\abs{\Gamma\backslash H_t}>\frac\epsilon 2$ holds.
If the latter holds, it should be rejected with high probability; hence we consider the former case.
Suppose that the oracle passes all homomorphism tests.
From Proposition \ref{prop}, since it passes the homomorphism tests the distance between $H_{t-1}$ and any solvable group $\tilde H_{t-1}$ is at least $\frac\epsilon 2$.
By repeating the above arguments, we see that the cyclic group $H_1$ is $\frac\epsilon 2$-far from any cyclic group but it passes the test for Abelian groups.
This is a contradiction.

We turn to the time complexity of the algorithm.
The number of iterations of the task 3 in the diagram in subsection 3.1, $t$, is polynomial because the output size of the task 1 is also polynomial.
We can easily know that the computation time of each task other than 3-(b) is polynomial.
How many times do we need to repeat the homomorphism tests?
Theorem \ref{hom} and Proposition \ref{prop} tell us that suppose the stage $j=k$ is the first stage such that the distance between $H_k$ and $\tilde H_k$ is greater than $\epsilon$ we fail in the homomorphism test with probability greater than $\eta$.
When we repeat the tests $c$ times, the probability never to fail is less than $(1-\eta)^c$.
We are able to take $c=O(\frac 1\eta)$ to bound the success probability by $\frac 1 3$.
As $\eta$ is taken to be a constant times $\epsilon$, the total computation time is polynomial in $\log\Gamma$ and $\epsilon^{-1}$ as we stated.


\begin{thebibliography}{000}

\bibitem{adpr}
N.~Alon, S.~Dar, M.~Parnas and D.~Ron, 
\emph{Testing of clustering}, 
Proceedings of the 41st Annual IEEE Symposium on Foundations of Computer Science, 240--250, 2000.
\bibitem{afns}
N.~Alon, E.~Fischer, I.~Newman and A.~Shapira, 
\emph{A combinatorial characterization of the testable graph properties: It's all about regularity}, 
Proceedings of the 37th Annual ACM Symposium on Theory of Computing, 251--260, 2006. 
\bibitem{anks}
N.~Alon, M.~Krivelevich, I.~Newman and M.~Szegedy, 
\emph{Regular languages are testable with a constant number of queries}, 
Proceedings of the 40th Annual IEEE Symposium on Foundations of Computer Science, 656--666, 1999.
\bibitem{babai}
L.~Babai, G.~Cooperman, L.~Finkelstein, E.~Luks and \'A.~Seress, 
\emph{Fast Monte Carlo algorithms for permutation groups}, 
Journal of Computer and System Sciences, 50:296--307, 1995.
\bibitem{bacon}
D.~Bacon, A.~Childs and W.~van Dam.
\emph{From optimal measurement to efficient quantum algorithms for the hidden subgroup problem over semidirect product groups}
Proceedings of the 46th Annual IEEE Symposium on Foundations of Computer Science, 469--478, 2005.
\bibitem{ben-or}
M.~Ben-Or, D.~Coppersmith, M.~Luby and R.~Rubinfeld, 
\emph{Non-Abelian homomorphism testing, and distributions close to their self-convolutions}, 
Proceedings of the 8th International Workshop on Randomization and Computation, 273--285, 2004.
\bibitem{blum}
M.~Blum, 
\emph{Designing programs to check their work}, 
Technical Report TR-88-009, International Computer Science Institute, 1988.
\bibitem{bk}
M.~Blum and S.~Kanna, 
\emph{Designing programs that check their work}, 
Proceedings of the 21st Annual ACM Symposium on Theory of Computing, 86--97, 1989.
\bibitem{blr}
M.~Blum, M.~Luby and R.~Rubinfeld, 
\emph{Self-testing/correcting with applications to numerical problems}, 
Proceedings of the 22nd Annual ACM Symposium on Theory of Computing, 73--83, 1990.
\bibitem{buhrman}
H.~Buhrman, L.~Fortnow, I.~Newman and H.~R\"ohrig, 
\emph{Quantum property testing},
Proceedings of the 14th annual ACM-SIAM Symposium on Discrete algorithms, 873--882, 2001.
\bibitem{czumaj00}
A.~Czumaj and C.~Sohler, 
\emph{Property testing in computational geometry}, 
Proceedings of the 8th Annual European Symposium on Algorithms, 155-166, 2000.
\bibitem{ergun}
F.~Erg\"un, S.~Kannan, R.~Kumar, R.~Rubinfeld and M.~Viswanathan, 
\emph{Spot-checkers}, 
Journal of Computer and System Sciences, 60(3):717--751, 2000.
\bibitem{fenner}
S.~Fenner and Y.~Zhang, 
\emph{Quantum algorithms for a set of group theoretic problems}, 
Proceedings of the 9th Italian Conference on Theoretical Computer Science, Lecture Notes in Computer Science, 3701:215--227, 2005.
\bibitem{fischer}
E.~Fischer, 
\emph{The art of uninformed decisions: A primer to property testing}, 
The Computational Complexity Column of The Bulletin of the European Association for Theoretical Computer Science, 75:97--126, 2001.
\bibitem{fkrss}
E.~Fischer, G.~Kindler, D.~Ron, S.~Safra and A.~Samorodnitsky, 
\emph{Testing juntas},
Proceedings of the 43rd Annual IEEE Symposium on Foundations of Computer Science, 103--112, 2002.
\bibitem{flnrrs}
E.~Fischer, E.~Lehman, I.~Newman, S.~Raskhodnikova, R.~Rubinfeld and A.~Samorodnitsky, 
\emph{Monotonicity testing over general poset domains}, 
Proceedings of the 34th Annual ACM Symposium on Theory of Computing, 474--483, 2002.
\bibitem{fimss}
K.~Friedl, G.~Ivanyos, F.~Magniez, M.~Santha and P.~Sen.
\emph{Hidden translation and orbit coset in quantum computing}, 
Proceedings of the 35th Annual ACM Symposium on Theory of Computing, 1--9, 2003. 
\bibitem{friedl}
K.~Friedl, G.~Ivanyos and M.~Santha, 
\emph{Efficient testing of groups}, 
Proceedings of the 37th Annual ACM Symposium on Theory of Computing, 157--166, 2005.
\bibitem{fmss}
K.~Friedl, F.~Magniez, M.~Santha and P.~Sen, 
\emph{Quantum testers for hidden group properties}, 
Proceedings of the 28th International Symposium on Mathematical Foundations of Computer Science, Lecture Notes in Computer Science, 2747:419--428, 2003.
\bibitem{goldreich}
O.~Goldreich, S.~Goldwasser and D.~Ron, 
\emph{Property testing and its connection to learning and approximation}, 
Proceedings of the 37th Annual IEEE Symposium on Foundations of Computer Science, 339-348, 1996.
\bibitem{hallgren}
S.~Hallgren, A.~Russel and A.~Ta-Shma.
\emph{Normal subgroup reconstruction and quantum computing using group representations}, 
Proceedings of the 32th Annual ACM Symposium on Theory of Computing, 627--635, 2000.
\bibitem{psi}
Y.~Inui and F.~Le Gall.
\emph{An efficient algorithm for the hidden subgroup problem over a class of semi-direct product groups}, 
arXiv: quant-ph/0412033, 2004.
\bibitem{kuperberg}
G.~Kuperberg,
\emph{A subexponential-time quantum algorithm for the dihedral hidden subgroup problem},
SIAM Journal on Computing, 35(1):170--188, 2005.
\bibitem{lipton}
R.~Lipton, 
\emph{New directions in testing}, 
DIMACS Series in Discrete Mathematics and Theoretical Computer Science, 2:191--202, 1991.
\bibitem{lomont}
C.~Lomont
\emph{The hidden subgroup problem - Review and open problems}, 
arXiv: quant-ph/0411037, 2004.
\bibitem{magniez}
F.~Magniez and A.~Nayak, 
\emph{Quantum complexity of testing group commutativity}, 
Proceedings of 32nd International Colloquium on Automata, Languages and Programming, Lecture Notes in Computer Science, 1770:1312--1324, 2005.
\bibitem{moore} 
C.~Moore, D.~Rockmore, A.~Russell and L.~Schulman. 
\emph{The power of basis selection in Fourier sampling: Hidden subgroup problems in affine groups}, 
Proceedings of the 15th Annual ACM Symposium on Discrete Algorithms, 1106--1115, 2004. 
\bibitem{rajagopalan}
S.~Rajagopalan and L.~Schulman, 
\emph{Verification of identities}, 
Proceedings of the 37th Annual IEEE Symposium on Foundations of Computer Science, 612--616, 1996.
\bibitem{regev}
O.~Regev,
\emph{Quantum computation and lattice problems},
SIAM Journal on Computing, 33(3), 738--760, 2004.
\bibitem{ron}
D.~Ron, 
\emph{Property testing}, 
In Handbook of Randomized Computing, Kluwer Academic Publishers, 597- -649, 2001.
\bibitem{rotteler}
M.~R\"{o}tteler and T.~Beth.
\emph{Polynomial-time solution to the hidden subgroup problem for a class of non-Abelian groups}, 
arXiv: quant-ph/9812070, 1998.
\bibitem{rubinfeld}
R.~Rubinfeld and M.~Sudan, 
\emph{Robust characterizations of polynomials with applications to program testing}, 
SIAM Journal of Computing, 25(2), 252--271, 1996.
\bibitem{shor}
P.~Shor, 
\emph{Polynomial-Time algorithms for prime factorization and discrete logarithms on a quantum computer}, 
SIAM Journal on Computing, 26(5), 1484--1509, 1997.
\bibitem{watrous}
J.~Watrous, 
\emph{Quantum algorithms for solvable groups}, 
Proceedings of the 33rd Annual ACM Symposium on Theory of Computing, 60--67, 2001.

\end{thebibliography}
\end{document}